\documentclass[iop]{emulateapj}
\usepackage{ifpdf}
\ifpdf
\usepackage[iop]{emulateapj}
\else

\fi

\slugcomment{To appear in AJ}

\shorttitle{Dust and Radio Jet in NGC~4258}
\shortauthors{Laine et al.}

\begin{document}

\title{Lack of Interaction between the Dust Grains and the Anomalous Radio Jet in the Nearby Spiral Galaxy NGC~4258}

\author{Seppo Laine}
\affil{{\it Spitzer} Science Center - Caltech, MS 220-6,
Pasadena, CA 91125; seppo@ipac.caltech.edu}

\author{Marita Krause}
\affil{Max-Planck-Institut f\"{u}r Radioastronomie, Auf dem
H\"{u}gel 69, 53121 Bonn, Germany; mkrause@mpifr-bonn.mpg.de}

\author{Fatemeh S. Tabatabaei}
\affil{Max-Planck-Institut f\"{u}r Radioastronomie, Auf dem
H\"{u}gel 69, 53121 Bonn, Germany; tabataba@mpifr-bonn.mpg.de}

\and

\author{Christos Siopis}
\affil{Institut d'Astronomie et d'Astrophysique, Universit\'{e} Libre de Bruxelles, CP 226, Boulevard du Triomphe, 1050 Bruxelles, Belgium; christos.siopis@ulb.ac.be}

\begin{abstract} 

We obtained {\it Spitzer}/IRAC 3.6 -- 8~$\mu$m images of the nearby spiral
galaxy NGC~4258 to study possible interactions between dust and the radio jet.
In our analysis we also included high-resolution radio continuum, H$\alpha$, CO,
and X-ray data. Our data reveal that the 8~$\mu$m emission, believed to
originate largely from PAH molecules and hot dust, is an excellent tracer of the
normal spiral structure in NGC~4258, and hence it originates from the galactic
plane. We investigated the possibility of dust destruction by the radio jet by
calculating correlation coefficients between the 8~$\mu$m and radio continuum
emissions along the jet in two independent ways, namely (i) from
wavelet-transformed maps of the original images at different spatial scales, and
(ii) from one-dimensional intensity cuts perpendicular to the projected path of
the radio jet on the sky. No definitive sign of a correlation (or
anticorrelation) was detected on relevant spatial scales with either approach,
implying that any dust destruction must take place at spatial scales that are
not resolved by our observations.

\end{abstract}


\section{INTRODUCTION}

The nearby (D=7.2~Mpc; \citeauthor{herrnstein99} \citeyear{herrnstein99}) spiral
galaxy NGC~4258 is known for its central black hole \citep{miyo95, herrnstein98}
and jets (anomalous arms), and is unique in its geometry among nearby spiral
galaxies. It possesses kpc-scale anomalous radio continuum and H$\alpha$ and
X-ray arms, in projection perpendicular to its normal spiral arms
\citep*{court61,kruit72} that have been shown to connect to a nuclear jet with
sub-arcsecond resolution observations \citep*{herrnstein97,cecil00}. The
H$\alpha$ emission along the jet is not excited by OB stars \citep{court61} but
instead is consistent with shock excitation \citep*[e.g.,][]{cecil95}. The
anomalous radio arms emerge as linear features from the nucleus up to a
projected distance of 840~pc, and then have a first bend of about $40\degr$,
followed by further bends, in the trailing sense relative to the galactic
rotation \citep{krause04}. 

The accretion disk of the central engine has an inclination angle of $83\degr$
and a position angle (p.a.) of $86\degr$ \citep{miyo95}, i.e., it is oriented
nearly east--west. The p.a. of the galactic disk, however, is $150\degr$, which
is nearly north-south, and its inclination is $72\degr$ \citep{albada80}. The
plane of the galactic disk and the plane of the accretion disk have
significantly differing orientations, being tilted by $60\degr$~--~$83\degr$
with respect to each other \citep*{krause07}. As the jets emerge perpendicular
to the accretion disk, they have to travel through the galactic disk at a rather
small angle. The distribution of the molecular clouds (see below) and the
H$\alpha$ emission of the jets indicate a direct interaction of the jets with
disk material within 2~kpc on either side of the nucleus.

Several studies \citep*[e.g.,][]{cecil92,cecil00} have elaborated on the ``jet''
structure of the anomalous radio continuum arms. One of the most comprehensive
studies in the radio continuum and optical wavelengths was made by
\citet{cecil00}. They find the jet to be in projection perpendicular to a
nuclear accretion disk (but in space misaligned by $65\degr$), and  associated
with two radio hot spots at around 1--2 kpc from the nucleus. They find optical
emission-line arcs on the leading edges of the radio hot spots. They explain the
observed structure with a shock that propagates with a velocity of 350$\pm$100
km~s$^{-1}$ but is now stalled. The shock connects to the anomalous radio
continuum arms. They interpret the overall structure with a precessing jet that
has recently moved out of the plane of the galaxy. The outer anomalous arms may
represent gas that has been driven out of the plane by the jet-affected halo gas
that is now impacting the disk, causing an apparent curvature when projected to
the plane of the sky \citep*{wilson01}.

Single-dish $^{12}\rm{CO}$ 1--0 line observations carried out with the 30-m IRAM
telescope revealed molecular gas in NGC\,4258 only in the center and along the
H$\alpha$ jets, at distances up to 2\,kpc \citep{krause90,cox96}. It was not
detected in other parts of the (central) galactic disk, down to the detection
limit of about 30\,mK brightness temperature. The velocities of the molecular
gas correspond primarily to the rotation of the galactic disk. Hence the
molecular clouds are in the plane of the disk. High-resolution
$^{12}\rm{CO}$~1--0 observations of \citet{krause07} with the IRAM
interferometer at Plateau de Bure revealed that the bulk of the CO emission
comes from two parallel ridges on both sides of the linear radio jets that cross
the nucleus. This morphology was confirmed by BIMA $^{12}\rm{CO}$~1--0 and
Sub-Millimeter Array (SMA) $^{12}\rm{CO}$~2--1 observations \citep{sawada07}.
\citet{krause07} proposed a model in which the molecular gas temporarily
accumulates along ridges surrounding the jet due to an interaction of the jet
magnetic field with the molecular gas clouds.

While star formation has presumably not (yet) been triggered by the radio jet,
it is of interest to study how the dust and gas have been affected by the
burrowing radio jet. Previous studies of galactic jet--gas or jet--dust
interactions  \citep[e.g.,][]{whittle04} suggest that molecular gas clouds that
cross the  path of the jet will cause bends and knots in the jet, but eventually
the jet may penetrate and ablate the cloud, in the process ionizing and
accelerating the gas. In radio galaxies the high-velocity jets ($v_{\rm jet}$ =
10$^{4}$ -- 10$^{5}$ km~s$^{-1}$) coming from the nuclei are expected to destroy
dust grains very quickly \citep{deyoung98}. In NGC~4258 the jet speed
is an order of magnitude smaller, around 1000 km~s$^{-1}$ \citep{cecil92}, and
therefore it may be insufficient to cause dust destruction. 

\begin{figure*}
\centering
\includegraphics[width=15cm]{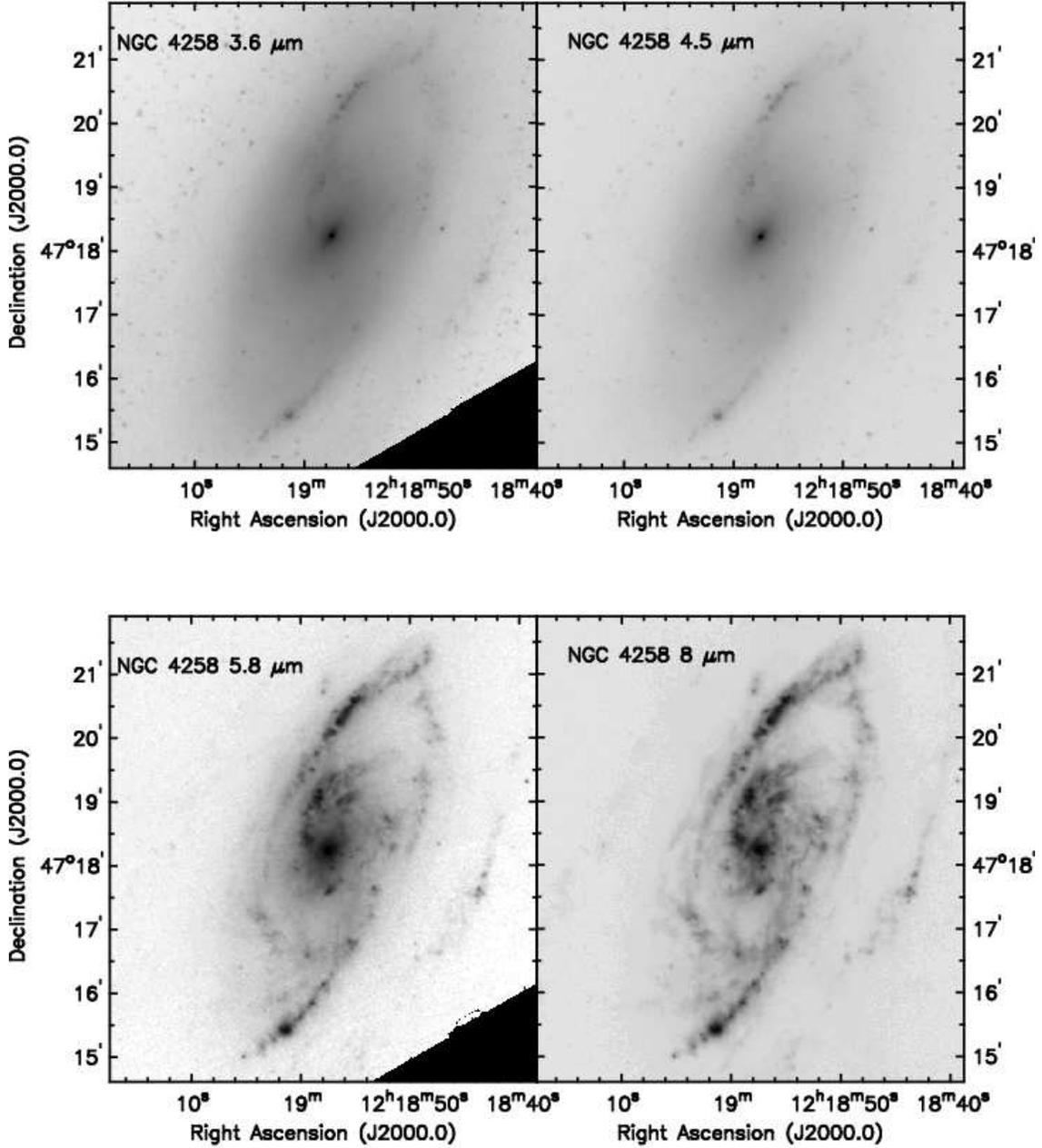}
\caption{Grayscale 3.6, 4.5, 5.8, and 8 $\mu$m IRAC images of NGC 4258.}
\label{mosa}
\end{figure*}

\begin{figure}
\centering
\includegraphics[width=8cm]{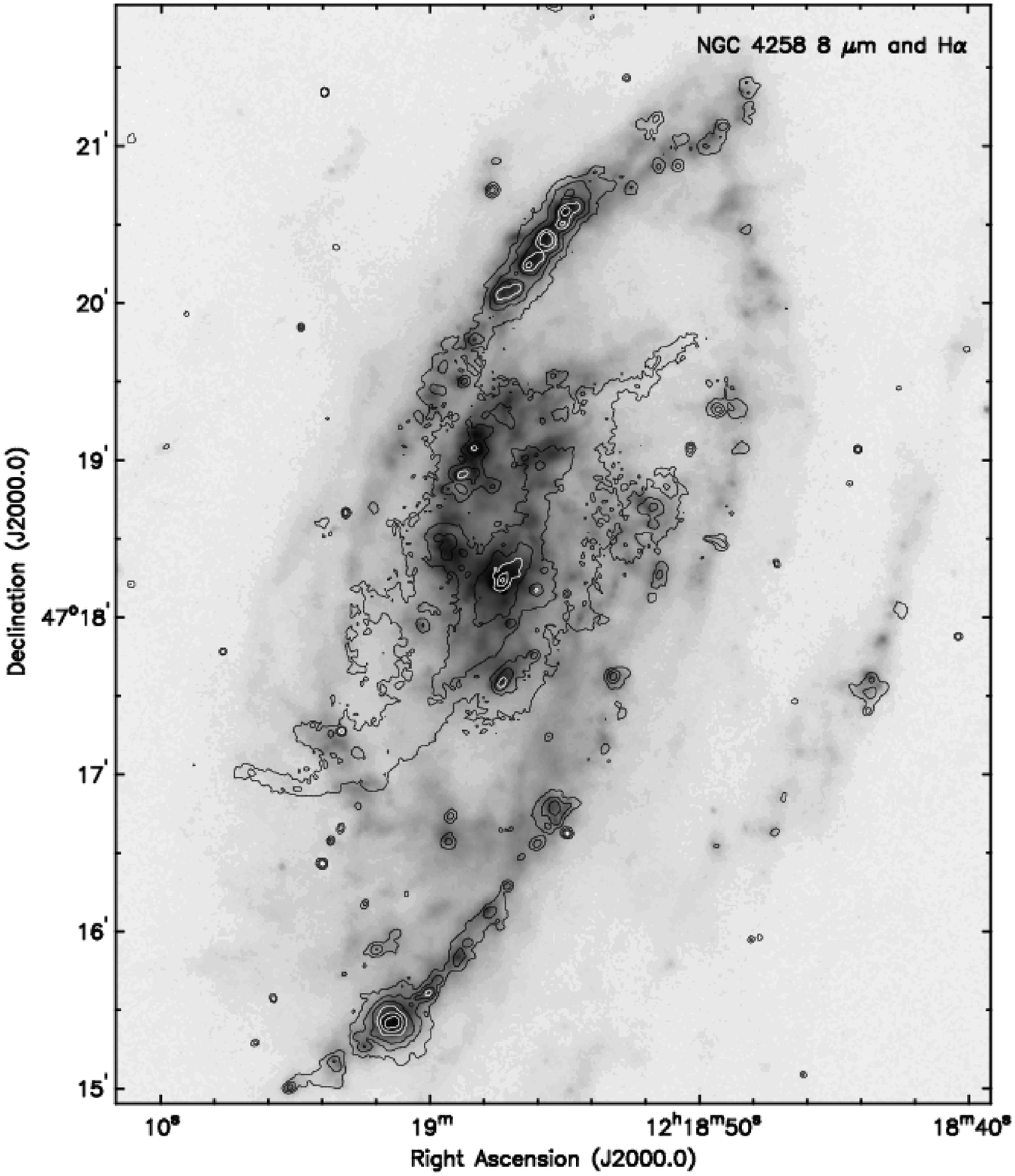}
\caption{Grayscale 8 $\mu$m image of NGC~4258 with overlaid contours of the H$\alpha$ emission from \citet{krause07}. The contours are in arbitray units
and logarithmically spaced at 10.0~$\times$~(2, 4, 8, 16, 32, 64, 128).}
\label{8-Halpha}
\end{figure}

We obtained {\it Spitzer}/Infrared Array Camera (IRAC) images of NGC~4258 from
3.6 to 8~$\mu$m. The 8~$\mu$m image is dominated by 7.7~$\mu$m and 8.6~$\mu$m
polycyclic aromatic hydrocarbon (PAH) emission features and radiation from hot
dust \citep[e.g.,][]{regan04}. Similarly, the 5.8~$\mu$m image is dominated by
the 6.2 $\mu$m PAH emission feature. We compare the IRAC images to images of the
radio continuum, CO, H$\alpha$, and X-ray emission at a comparable resolution to
investigate how the radio jet may have affected the dust properties. We use  the
8~$\mu$m emission as a tracer of dust emission, the 3.6 cm radio continuum and
X-ray emission as a tracer of the jet, and we look for signs of possible dust
destruction by the jet by searching for an anticorrelation between these
emissions. We do this with the help of a wavelet analysis and by looking at
one-dimensional intensity slices taken across the projected position of the jet
in the images.

\section{OBSERVATIONS}

NGC~4258 was observed with IRAC \citep{fazio04} aboard the {\it Spitzer Space
Telescope} on 2005 December 24. Observations with the channel 2 (4.5 $\mu$m) and
4 (8~$\mu$m) cameras were centered on the galaxy. Observations in channels 1
(3.6 $\mu$m) and 3 (5.8 $\mu$m) also covered the central part of the galaxy. The
total mapped area was about 28\arcmin\ $\times$ 26\arcmin\ in each channel. We
used the small-scale cycling dithering option together with half array map steps
in the array column and row directions to cover the required area on the sky and
to produce sufficient redundancy against data artifacts. We also used the
high-dynamic-range observing mode which takes a short-exposure image before a
long-exposure image. The short-exposure frame times were 1.2 seconds and the
long-exposure frame times were 30 seconds. In this paper we use only the
30-second frames, as we did not find any saturated areas in the galaxy. The
pixel size in each channel is roughly 1\farcs 2\ $\times$ 1\farcs 2. The
observations reported here use the S14 pipeline processing version. We ran the
individual IRAC basic calibrated data (BCD) frames through the IRAC artifact
mitigation software (available as contributed software on the {\it Spitzer}
website). Subsequently we improved the muxbleed artifact mitigation in the
long-exposure images from channels 1 and 2 by custom software, kindly provided
by J. Hora. We then made mosaics of the long-exposure frames with the Mosaicking
and Point Source Extraction (MOPEX) mosaicking software \citep{makovoz05},
available at the {\it Spitzer} Science Center website. We corrected for the
background variations in the individual frames by running the BCDs through the
overlap correction algorithm first. This algorithm normalizes the spatially
overlapping parts of individual BCD frames to the same mean value. The final
pixel size in the mosaics is 0\farcs6\ $\times$ 0\farcs 6. MOPEX rejects
outliers, such as radiation hits, during the construction of the mosaic. The
standard interpolation method was used, whereby the relative overlapping areas
of the input pixels from the individual BCDs are used as weights for calculating
the value of the output pixel.

\begin{figure}
\centering
\includegraphics[width=8cm]{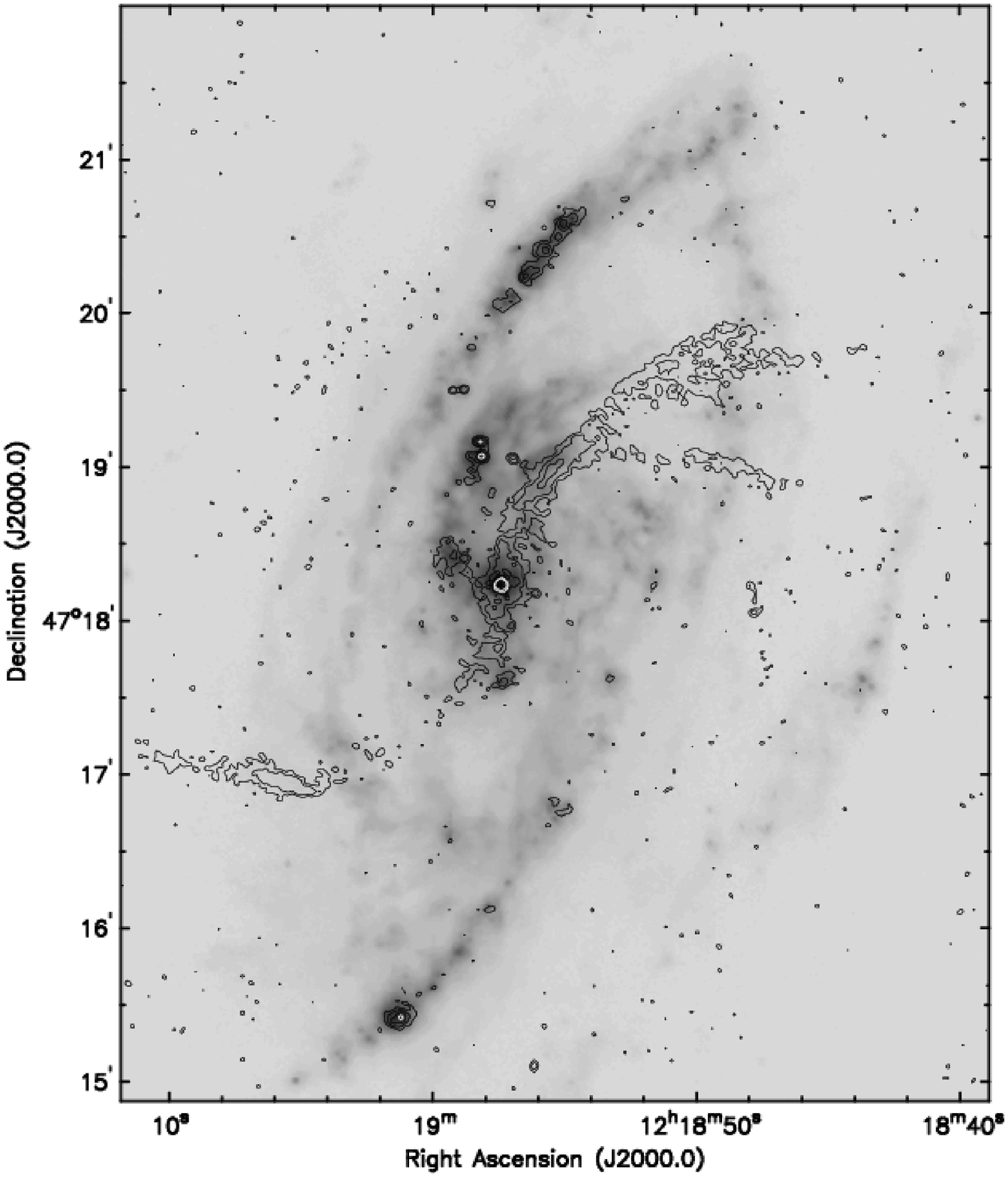}
\caption{Grayscale 8 $\mu$m image of NGC~4258 with overlaid contours of the high resolution radio continuum emission at 8.44~GHz (HPBW~=~$2\farcs2 \times
2\farcs4$) from \citeauthor{krause04} \citeyear{krause04}. The contours are at 6.0~$\times$~10$^{-6}$ $\times$ (4, 6, 8, 16, 32, 64) Jy~beam$^{-1}$.
\label{8-radio}}
\end{figure}

\section{RESULTS}

\subsection{Morphology}

The final IRAC mosaics are shown in Figure~\ref{mosa}. The emission at
3.6~$\mu$m and 4.5~$\mu$m is very smooth over most of the galaxy, and traces the
older stellar population in the disk. The strongest star formation regions in
the spiral arms stand out from the smooth overall distribution of the emission
at these wavelengths. There is likely to be a contribution from bright
supergiant stars and possibly from hot dust in regions of strong recent or
active star formation. The active nucleus of NGC~4258 shows up as a bright,
point-like component, which is likely to have a larger contribution from hot
dust as the wavelength increases from 3.6 $\mu$m to 8 $\mu$m. The morphology
becomes patchier in the 5.8 and 8 $\mu$m images. At 5.8~$\mu$m, the stellar
contribution decreases dramatically and the 6.2~$\mu$m PAH and hot dust
continuum components become stronger. The 8~$\mu$m IRAC band still traces hot
dust, but the PAH emission is likely to dominate. The hot dust and PAH emissions
appear to trace the spiral arms very well (as discussed in the next section) and
are found around star formation regions, but the anomalous H$\alpha$ and radio
continuum arms have very little hot dust or PAH emission. A similar change in
morphology as the wavelength increases from 3.6 $\mu$m to 8 $\mu$m has been seen
before, e.g., in M81 \citep{willner04}.

\begin{figure}
\centering
\includegraphics[width=8cm]{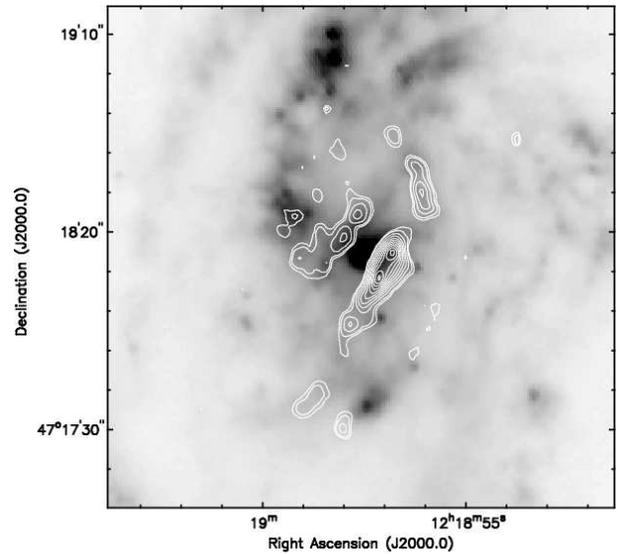}
\caption{CO emission shown with contours on a grayscale image of the 8 $\mu$m emission. The contours are shown at 2.0~$\times$~(3, 4, 6, 8, 10, 12, 14, 16, 18, 20, 22, 24) Jy~beam$^{-1}$~km~s$^{-1}$. The radio jet position angle is $-3\degr$ for $r$~$<$~24\arcsec, and $-43$\degr\ for $r$~$>$~24\arcsec, whereas the CO and H$\alpha$ (and X-ray) emission are aligned at about $-25\degr$.
\label{8-co}}
\end{figure} 

We also measured the ellipticity in the 3.6 $\mu$m and 4.5 $\mu$m images at
radii within the major star forming spiral arms. The ellipticity rises
to a value  of around 0.35--0.4 at semimajor axes from $\sim$~3\arcsec\ to
$\sim$ 30\arcsec\ (100 pc -- 1 kpc). The position angle is around 146\degr\ --
148\degr. This is very similar to the 150\degr\ position angle and 72\degr\
inclination of the galactic disk of NGC~4258, as determined from \ion{H}{1}
observations by \citet{albada80}. We do not see any evidence for a lense-like
structure (bar) at a position angle of $\sim$~10\degr\ (contrary to what was
found by \citeauthor{albada80} \citeyear{albada80} in \ion{H}{1}) in our images, but we cannot rule out the existence of an oval distortion either.

We do not see the radio continuum hot spots (about 50 arcseconds north and 25
arcseconds south of the nucleus) in the 8~$\mu$m emission. They are located in
relative low intensity valleys of the 8 micron emission. The predicted flux
densities of the hot spots, as extrapolated from the radio continuum and
assuming a spectral index of 0.7, would be less than 1~$\mu$Jy, well below
the extended dust continuum background radiation from the disk which has fluxes
about 2.5 mJy at the locations of the radio hot spots (as integrated in
5\arcsec\ $\times$ 5\arcsec\ boxes around the hot spots). 

\subsection{8-micron Emission as a Tracer of Spiral Structure}

We compared the maps at 5.8~$\mu$m and 8~$\mu$m with various other
high-resolution observations of NGC~4258 that were available to us, such as the
3.6-cm radio continuum \citep{krause04}, H$\alpha$ \citep{krause07}, X-ray data
from \citet{yang07}, and $^{12}\rm{CO}$~1--0 line observations \citep{krause07}.
We show the most interesting overlays of 8~$\mu$m with H$\alpha$, and 8~$\mu$m
with 3.6-cm radio continuum in Figures~\ref{8-Halpha} and \ref{8-radio},
respectively. As Figure~\ref{8-Halpha} shows, there is a very good spatial
correspondence between the 8 $\mu$m and H$\alpha$ emission regions along the
normal spiral arms of NGC~4258. However, there is much less spatial
correspondence in the central part of the galaxy and along the jets. The spatial
correspondence of the 8~$\mu$m emission with the 3.6-cm radio continuum emission
(Fig.~\ref{8-radio}) is much weaker. They only coincide around the nucleus and
along the star-forming areas of the normal spiral arms. It is known that the
radio continuum emission outside the normal star forming spiral arms is mainly
synchrotron emission from the jets \citep*[e.g.,][]{hummel89}, whereas the
H$\alpha$ emission traces both the normal spiral structure and the jets
\citep[e.g.,][]{cecil92}. The X-ray emission morphology is similar to the  radio
continuum morphology, although in the central two arcminutes the jet-like
feature in the X-ray image is almost perfectly straight, as can be seen in
Fig.~1 of \citet{yang07}. At larger distances the bulk of the x-ray emission
follows the radio continuum very closely, although the relative intensities at
various positions differ. The H$\alpha$ emission along the jet  follows the
X-ray emission closely. It has been suggested \citep{wilson01} that the jet is
inclined by about $30\degr$ to the disk of the galaxy and exits the disk at a
distance of a few hundred pc from the nucleus. Beyond the nucleus the linear
section of the X-ray jet could be due to shocks, triggered by the
out-of-the-plane radio continuum jet, impacting the disk plane and heating the
gas there. The curvature of the X-ray and radio continuum jets beyond a few kpc
from the nucleus could then be due to the lower density shock-heated gas that
has been pushed out of the galactic plane (to the other side of the galactic
plane with respect to the radio jet), thus causing the curvature, as seen in
projection \citep{wilson01}.

Quite a different spatial distribution is visible in $^{12}\rm{CO}$~1--0 line
observations when compared to the 8~$\mu$m emission (Figure~\ref{8-co}). The CO
emission is aligned along the jets and  concentrated within the innermost 2~kpc
in radius, unlike what is seen  in the 8~$\mu$m dust emission. The alignment of
the CO emission on the sides of the jet was interpreted as the molecular clouds
of the disk interacting with the magnetic field of the jets \citep{krause07}.
Such an interaction is, however, improbable for hot dust or PAHs.

We conclude that the 8~$\mu$m emission, emitted mainly by PAH dust
features and hot dust (which are evidently components of the galactic disk), is
an excellent tracer of the normal spiral structure of NGC~4258. This is
supported by the wavelet analysis of the various emissions, described in the following section.

\subsection{Wavelet Analysis of the 8-micron, Radio Continuum, And 
H$\alpha$ Components in NGC 4258}
\label{sec:wavelet}

\begin{figure*}
\centering
\includegraphics[width=15cm]{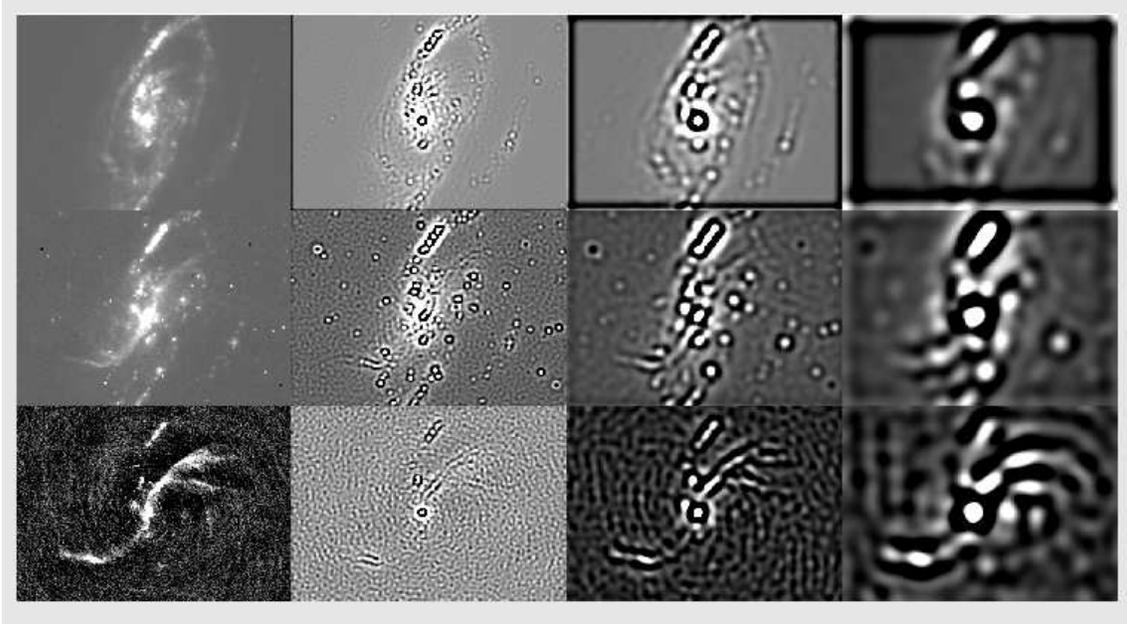}
\caption{From top to bottom: IRAC 8~$\mu$m, H$\alpha$, and 3.6\,cm radio continuum maps at 2\farcs 5~resolution (first column) and their wavelet-decomposed maps on a scale of 7\farcs 7~(second column),  16\arcsec~(third column), and 35\farcs 5~(fourth column). At the distance to NGC~4258, 1\arcsec\ $\approx$~35\,pc. The maps are shown in the north up, east left orientation and are centered on the nucleus of the galaxy.\label{multiplot}}
\end{figure*}

\begin{figure}
\centering
\includegraphics[width=8cm]{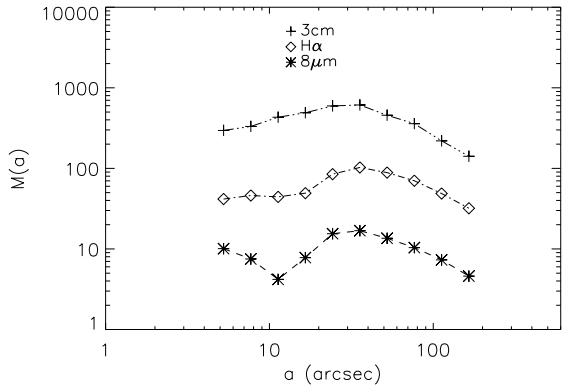}
\caption{Wavelet power spectrum for the 8\,$\mu$m, H$\alpha$, and 3.6 cm emissions along the track covered by the radio continuum jet. The data points correspond to the scales of 5.2, 7.7, 11.3, 16.6, 24.3, 35.6, 52.3, 67.7, 112.5, and 165.0 arcseconds. The power is shown in arbitrary units.
\label{spect.jet}}
\end{figure}

To investigate the origin of the 8~$\mu$m emission and its connection to the
radio and H$\alpha$ emissions, we compared the emitting structures on different
spatial scales using a wavelet transformation. This allows us to detect
systematic correlations or anti-correlations on each spatial scale. 

The wavelet technique allows us to analyze the power emitted at different
spatial scales (through the wavelet spectrum), and is more robust against noise
than Fourier filtering \citep{frick01}. Cross-correlating the wavelet spectra
from each wavelength lets us compare the emission structures in the various
images systematically as a function of the spatial scale. We apply a
two-dimensional wavelet transformation to separate the diffuse emission
components from compact structures in the 8 $\mu$m, 3.6\,cm radio continuum, and
H$\alpha$ images. The wavelet coefficients of a 2D continuous wavelet
transformation are given by

\begin{equation}
W(a,{\bf x})=\frac{1}{a^{\kappa}} \int_{-\infty}^{+\infty}  f(\bf x')\psi^{\ast}(\frac{{\bf x'-x}}{{\it a}}){\it d}{\bf x'},
\label{eq:wave1}
\end{equation}
 
\noindent
where $\psi({\bf x})$ is the analyzing wavelet, ${\bf x} = (x,y)$ positional coordinates, $f({\bf x})$ is a two--dimensional function (the image being analyzed), and {\it a} and $\kappa$ are the scale and normalization parameters, respectively, (the ${\ast}$-symbol denotes the complex conjugate).
Following \citet{frick01} and \cite{taba07}, we use the ``Pet--Hat'' wavelet to obtain a good separation of scales and to find the scale of the dominant structures in the galaxy. The ``Pet--Hat'' wavelet is defined in Fourier space by the formula:

\begin{equation}
\psi(\bf k)=\left \{ \begin{array}{ll}
{\cos^2\Big(\frac{\pi}{2}\,\log_{2}\,(\frac{k}{2\pi})\Big)}  &
\,\,\,\,\,\pi \le k  \le 4\pi \\
0 &   k  < \pi  \, \,\, {\rm or} \,\,\,  k  > 4\pi , %
\end{array} \right.
\label{eq:wave2}
\end{equation}

\noindent
where ${\bf k}$ is the wave vector and $k=\vert \bf k \vert$.
The above transformation decomposes an image into ``maps'' of different scale. In each map, structures with the chosen spatial scale are prominent as they have larger coefficients than those with smaller or larger spatial scales. 

\begin{figure}
\centering
\includegraphics[width=8cm]{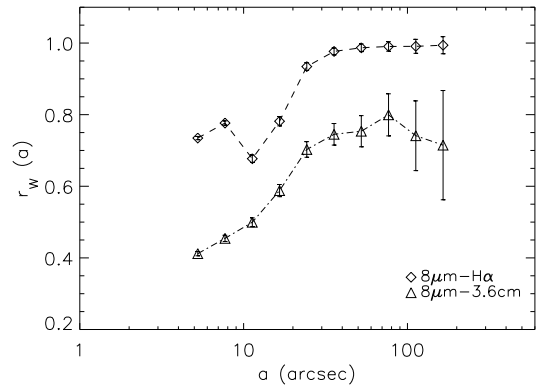}
\caption{Wavelet cross-correlation of the 8 $\mu$m emission with the H$\alpha$ and 3.6\,cm radio continuum emissions along the track covered by the radio continuum jet, as a function of spatial scale.\label{correl}}
\end{figure}

After converting the 8\,$\mu$m, 3.6\,cm, and H$\alpha$ images into the same
geometry, size, and resolution (2\farcs 5), we decomposed them into 10 spatial
scales ranging from 5\arcsec\ to 165\arcsec\footnote{To have both physically
meaningful results and a sufficiently large number of independent points, the
scale {\it a} varies between a minimum of about twice the resolution and a
maximum of about half of the image size.} using Equations~\ref{eq:wave1} and
\ref{eq:wave2}. Figure~\ref{multiplot} shows the original 8\,$\mu$m, 3.6\,cm,
and H$\alpha$ images together with their decomposed maps on three scales. On the
scale of $a$~=~7.7\arcsec\ (270\,pc), the  8\,$\mu$m emission reveals small
clumps that make up segments of rings or spiral arms, even where the H$\alpha$
emission is faint or absent. Furthermore, in contrast to the H$\alpha$ and radio
continuum emissions, there are filamentary and flocculent structures at
8\,$\mu$m, some of which appear more extended in the northern than in the
southern half of the galaxy. The outer spiral arms are better visible on the
$a$~=~16.5\arcsec\ (580\,pc) scale. At this scale the strong star-forming region
in the northern arm is the only common structure with the radio emission. On the
next larger scale, $a$~=~35.5\arcsec\ (1.2\,kpc), the dominant dust emitting
structures are found in the center and around the strong star-forming region
located in the northern spiral arm.

Apart from the center the radio continuum emission is dominated by the jet and
the star-forming region in the northern spiral arm on all scales. A large amount
of the H$\alpha$ emission appears as dispersed clumps of star-forming regions.
The H$\alpha$ emission jet appears as a smooth, elongated structure on scales
larger than 0.5\,kpc. However, there is not any trace of a jet-like structure in
the 8\,$\mu$m wavelet-decomposed maps. The 8\,$\mu$m emission traces the stellar
spiral arms much better than the H$\alpha$ or radio continuum emissions.

To quantify any correlation or anti-correlation of the dust emission with the
H$\alpha$ and radio continuum emissions along the jet, we focus on the region
covered by the radio jet. We made maps of the radio continuum, H$\alpha$, and
8\,$\mu$m emissions by including only regions corresponding to the radio
continuum intensities larger than a threshold (the point sources and the stellar
spiral arms had been subtracted from the radio continuum map before doing this).
These maps were then decomposed in the same way for the whole galaxy. In
Figure~\ref{spect.jet} we plot the wavelet spectrum, i.e., the distribution of
emitting power in terms of the spatial scale for each map given by 

\begin{equation}
M(a) = \int_{-\infty}^{+\infty} \int_{-\infty}^{+\infty}  \vert W(a,{\bf x})\vert ^{2} d\bf x.
\label{eq:spectrum}
\end{equation}

A smooth distribution of the power spectrum is expected if the emission
along the masked radio jet originates from a real jet. The radio emitting power increases toward the 20\arcsec\ -- 40\arcsec\
(0.7 -- 1.4\,kpc) scales and decreases smoothly beyond these scales. There does
not appear to be any overall dominant power at small spatial scales. However,
the smallest analyzed spatial scale (5\arcsec\ or 170\,pc) has the most power
among the first four analyzed spatial scales at 8\,$\mu$m. This indicates
patchiness of the dust emission along the radio jet region. These emission
patches probably emerge from gas and dust clouds in the disk. It is notable that
all three wavelengths share the same broad peak in their wavelet power
spectrum  between 20\arcsec\ and 40\arcsec. This reflects mostly the enhanced
emission (and size of the) central region.

The wavelet spectrum of the H$\alpha$ emission is intermediate in nature between
those of the radio and dust emissions. It is not dominated by the small scale
emission, in contrast to the 8\,$\mu$m emission, but it also does not increase
smoothly toward the maxima at 0.7--1.4\,kpc scales (unlike the radio continuum
power spectrum). H$\alpha$ shares the smooth decline of the other two power
spectra past the peak. This may be a consequence of the H$\alpha$ emission
consisting of both emission from the H$\alpha$ jet and the underlying disk
structures, specifically, \ion{H}{2} regions.

We calculate the correlation coefficients on each scale with
\begin{equation}
r_{\rm w}(a)=\frac{\int \int W_{1}(a,{\bf x})~W^{\ast}_{2}(a,{\bf x})~d{\bf x}}{[M_{1}(a)M_{2}(a)]^{1/2}},
\end{equation}

\noindent
where the subscripts refer to two images of the same size and linear resolution. 
The value of $r_{\rm w}$ varies between $-1$ (total anticorrelation) and +1 (total correlation).

The uncertainty in $r_{\rm w}$ can be approximated by

\begin{equation}
\Delta r_{\rm w}(a)= \frac{\sqrt{1-r^{2}_{w}}}{\sqrt{n-2}},
\end{equation}
 
\noindent where $n = 2.13 (L/a)^2$ is the  number of independent points and $L$
is the size of the maps. Plotting $r_{\rm w}$ against the spatial scale shows
how well structures at different scales are correlated in intensity and
location. The wavelet cross-correlation coefficients $r_{w}$ are shown in
Figure~\ref{correl}. The dust emission is not correlated with the radio emission
at small scales. A significant 8\,$\mu$m -- 3.6\,cm correlation ($r_{\rm
w}>0.75$;\footnote{A coefficient $r_{\rm w}$ of $>$ 0.75 means that more than
50\% of the total variation in one of the variables can be explained by a linear
relationship between the two variables.} although with large uncertainty), is
found at large scales ($a>36\arcsec$) due to the extended emission from the
bright central part and because of the artificial smearing of the two emissions
by using the same masking (at 2$\sigma$ radio continuum rms level, covering the
jet) in both maps. In general, the dust emission has a better correlation with
the H$\alpha$ emission (even in the radio jet region) than with the radio
continuum emission. The weakest correlation occurs around $a\sim$11\arcsec\, at
about the width of the H$\alpha$ jet. A good small spatial scale 8\,$\mu$m --
H$\alpha$ correlation is possibly due to the star-forming regions in the disk
that are projected on the radio-jet along the line of sight. If so, then a
superposition of the emission from the jet and the underlying galaxy disk makes
it very difficult to detect an anticorrelation between the dust and the radio
emissions along the jet.

\begin{figure*}
\centering
\includegraphics[width=15cm]{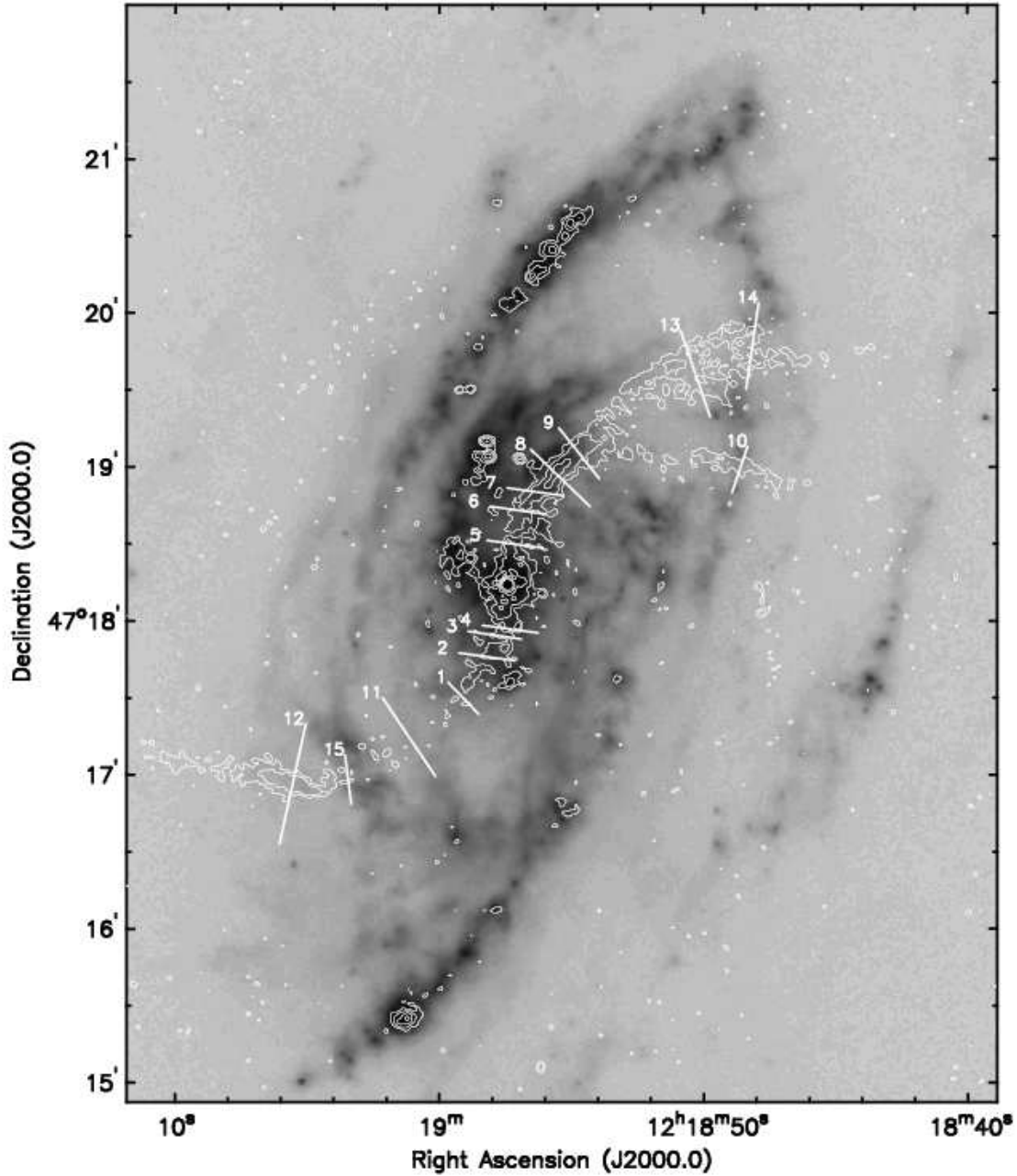}
\caption{Same as Fig.~\ref{8-radio} but with the labeled positions of the slices taken across the radio jet and shown in Figure~\ref{slice}.
\label{radio}}
\end{figure*}

\begin{figure*}
\centering
\includegraphics[width=15cm]{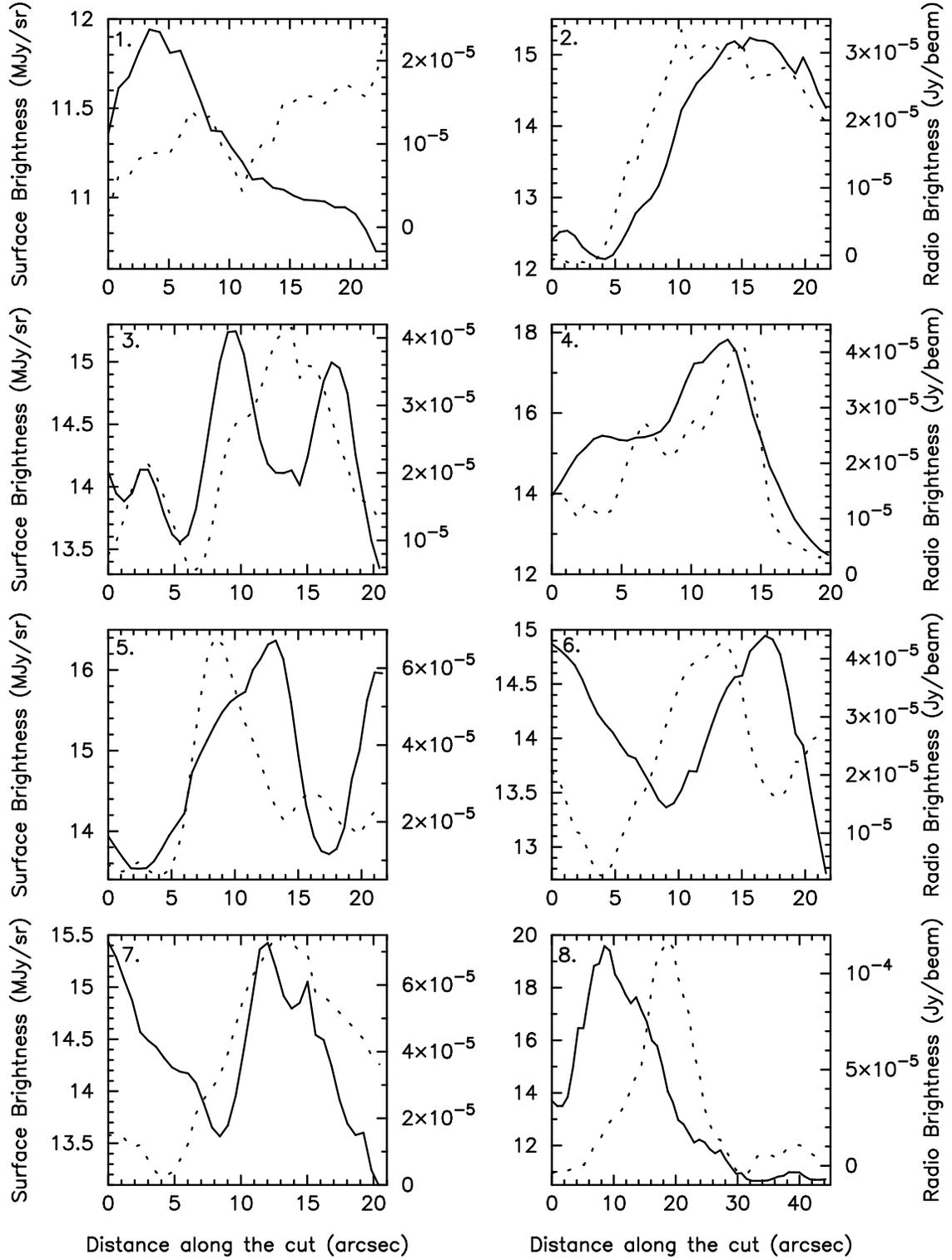}
\caption{Slices taken across the radio jet as shown in Fig.~\ref{radio}, in 8 $\mu$m (solid line) and 3.6 cm radio emission (dashed line).
\label{slice}}
\end{figure*}

\setcounter{figure}{8}

\begin{figure*}
\centering
\includegraphics[width=15cm]{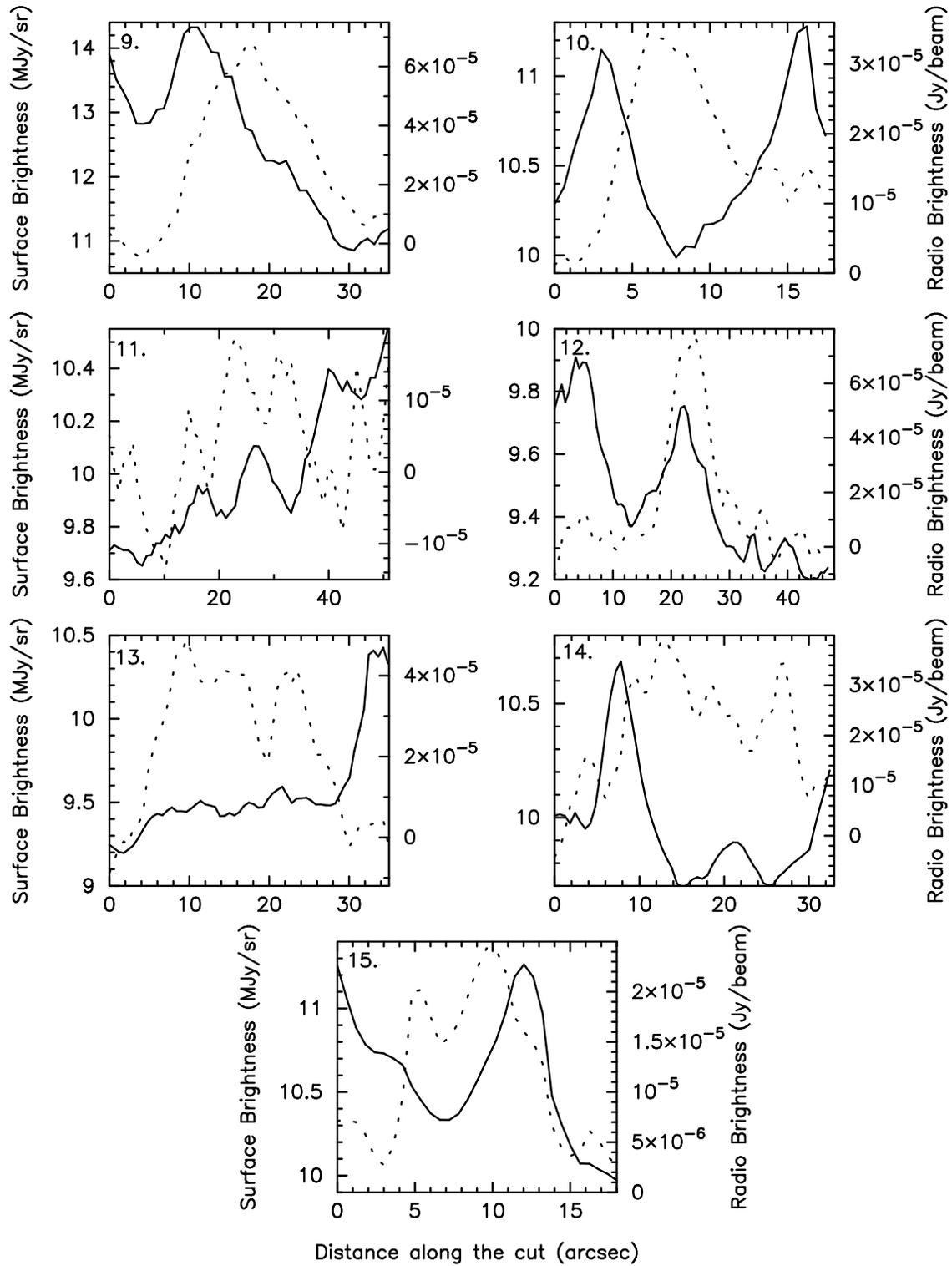}
\caption{Continued.}
\end{figure*}

\begin{deluxetable*}{lcccc}
\tablecaption{Correlation coefficients between the 8 $\mu$m and radio continuum emissions along the slices shown in Figure~\ref{slice}. \label{SliceCorrTable}}
\tablehead{
\colhead{} & \multicolumn{2}{c}{Pearson} & \multicolumn{2}{c}{Spearman} \\
\colhead{Slice} & \multicolumn{2}{c}{Correlation} & \multicolumn{2}{c}{Correlation} \\
\colhead{} & \colhead{Coefficient} & \colhead{$P$-value} & \colhead{Coefficient} & \colhead{$P$-value} }
\startdata
1&  -0.68 & $4.3 \times 10^{-5}$ & -0.75 & $2.3 \times 10^{-6}$ \\
2& 0.88  & $1.2 \times 10^{-12}$ & 0.75  & $1.1 \times 10^{-7}$ \\
3& 0.41  & $1.5 \times 10^{-2}$ & 0.49  & $2.8 \times 10^{-3}$ \\
4& 0.89  & $1.7 \times 10^{-13}$ & 0.91  & $9.9 \times 10^{-15}$ \\
5&  0.55  & $3.8 \times 10^{-4}$ & 0.61  & $7.2 \times 10^{-5}$ \\
6& -0.22 & $1.8 \times 10^{-1}$ & -0.27 & $1.0 \times 10^{-1}$ \\
7& 0.22  & $2.0 \times 10^{-1}$ & 0.25  & $1.4 \times 10^{-1}$ \\
8&  0.25  & $6.8 \times 10^{-2}$ & 0.41  & $2.5 \times 10^{-3}$ \\
9&  0.16  & $3.0 \times 10^{-1}$ & 0.08  & $6.1 \times 10^{-1}$ \\
10&-0.58 & $7.4 \times 10^{-4}$ & -0.60 & $4.2 \times 10^{-4}$ \\
11& 0.17  & $1.9 \times 10^{-1}$ & 0.22  & $9.0 \times 10^{-2}$ \\
12& 0.32  & $4.0 \times 10^{-3}$ & 0.35  & $1.8 \times 10^{-3}$ \\
13& -0.27 & $3.7 \times 10^{-2}$ & -0.08 & $5.4 \times 10^{-1}$ \\
14& -0.32 & $1.8 \times 10^{-2}$ & -0.44 & $6.8 \times 10^{-4}$ \\
15& 0.30  & $1.0 \times 10^{-1}$ & 0.35  & $5.4 \times 10^{-2}$ \\
\enddata
\end{deluxetable*}

The wavelet analysis confirms that the 8\,$\mu$m emission traces the normal stellar spiral arms much better than the radio continuum and H$\alpha$ emissions. The long continuous spiral arm in the northeast stands out, as does the split stellar spiral arm in the south and a long section of a spiral arm in the northwest. It is also clear that the emission just north (northeast) of the nucleus at declination $47\degr$19\arcmin\ comes from a spiral arm rather than from the jet. In this region no CO line emission has been detected, which is consistent with the result that the CO emission is only found flanking the jet \citep[see][]{krause90}. Even more notably, the wavelet analysis shows that the 8\,$\mu$m emission does not trace the radio continuum jet emission at all.

\subsection{Emission Profiles Across the Jet}
\label{subsec:emprofs}

In this section we further examine the dust to radio continuum correlation by
taking one-dimensional slices in the 8 $\mu$m and 3.6-cm radio continuum
emissions perpendicular to the sky-projected orientation of the radio jet at
several positions across the galaxy. The positions of the slices were chosen to
sample the radio jet at various positions along its length. The slice positions 
are shown in Figure~\ref{radio}, and the emission profiles along the slices in
Figure~\ref{slice}. The only slice in which visual inspection shows evidence of
an anticorrelation between the radio continuum and 8 $\mu$m emissions is at
position 10, which lies far out in the outer spiral arms, where the radio
continuum jet has split into several branches. We believe that this is the
result of a fortuitous distribution of the two emissions at this location,
rather than of any physical PAH grain destruction by the radio jet. In several
other slices we see a phase offset between the radio continuum and 8 $\mu$m
emissions, but there does not appear to be any clear pattern in the  magnitude
of the offset, in the leading or trailing sense, as a function of the distance
from the nucleus of the galaxy or the side of the galaxy (and jet). We varied
the orientation and the length of the slices in a few cases and confirmed that
the lack of any correlation or anticorrelation in the slices is not due to the
orientation or length of the slices (see also the note below about performing
the correlation coefficient calculation by excluding points at the ends of the
slices).

We can quantify the preceding remarks, obtained from visual inspection, by calculating correlation coefficients between the 8~$\mu$m and the radio continuum emission profiles for each slice presented in Fig.~\ref{slice}. We calculated the parametric Pearson (linear) correlation and the non-parametric Spearman (rank-order) correlation, along with their $P$-values, for all the slices. In principle, the Spearman correlation is more appropriate since there is no reason to expect that the emission profiles are drawn from normally distributed populations. In practice, both methods produced similar results, which are shown in Table~\ref{SliceCorrTable}. The $P$-value is the probability of obtaining the calculated correlation coefficient assuming the null hypothesis, i.e., that the correlation coefficient is in fact zero. If this probability is lower than a conventional value, often taken to be 0.05, the correlation coefficient is considered to be statistically significant.

Using a threshold of $\pm$0.7 for a correlation or anticorrelation, respectively, we find only 2--3 slice profiles out of 15 that fulfill this criterion. Changing the threshold to $\pm$0.6 does not increase the number of possible correlated or anticorrelated slice profiles by more than 1. Slice 4 has the most correlated profiles, and they are {\it positively} correlated. The next most correlated slices are 1 (negatively) and 2 (positively), followed by slices 5 (positively) and 10 (negatively). There is one more positively than negatively correlated slice, but this is not a statistically significant result. 

We note that it is possible to obtain higher correlation numbers by excluding
points at the ends of the slices. However, when doing so, we still do not find
strong evidence of an anticorrelation (or correlation) anywhere else except at
slice position 10.

We repeated this correlation analysis for the 8~$\mu$m and X-ray images and
found no evidence of an anticorrelation (or correlation) between those two
emissions. The only position where we see a significant anticorrelation is a
little northwest of slice position 9 in Figure~\ref{radio}, where the radio
continuum emission is relatively weak (but X-ray emission has a stronger ridge).
However, this ``anticorrelation'' is due to the one-sided distribution (on the
northeastern side only) of the 8~$\mu$m emission with respect to the X-ray jet,
and therefore does not provide evidence for any destruction of PAH molecules and
dust in the jet path.

\section{DISCUSSION AND CONCLUSIONS}

Our data reveal that the 8~$\mu$m emission (and the 5.8 $\mu$m emission),
believed to originate largely from PAH dust features and hot dust, is an
excellent tracer of the normal spiral structure in NGC~4258, and hence it
originates from the galactic plane. It allows us to distinguish between the
normal spiral arms and the jet, which is not trivial, especially in the northern
part. The other well-known spiral arm tracer, the H$\alpha$ emission, cannot be
used to distinguish the spiral arm and jet emissions in NGC~4258 as we find
H$\alpha$ emission (shock ionized) also along the inner jets.

We searched for a possible signature of dust destruction by the large-scale
jets by comparing the 8-micron emission to radio continuum emission in two
different ways. A wavelet analysis shows that the 8~$\mu$m emission exhibits a patchy structure, even along the large-scale jet, unlike the emission seen
in the radio continuum (or H$\alpha$). All three studied emissions (8~$\mu$m, 
H$\alpha$, and radio continuum) show largest wavelet powers at around 30\arcsec\ ($\sim$~1 kpc) scales, mostly due to the central region. A correlation analysis
of the wavelength-transformed emission at various scales shows that the 8 $\mu$m emission is not correlated (or anticorrelated) with the radio continuum emission
on any scales. Overall, the wavelet analysis shows that the 8~$\mu$m emission
traces the normal spiral arms much better than the radio continuum or H$\alpha$ emissions.
  
We also took one-dimensional intensity profile cuts perpendicular to the  radio
continuum jet at several arbitrary positions along the jet. We only detected an
anticorrelation between the 8~$\mu$m and 3.6-cm radio continuum emissions far
out in the disk where the jet has several branches. We consider this
anticorrelation to be a fortuitous superposition of the two emissions rather
than evidence for dust destruction. The profiles at smaller radii along the jet
sometimes show a phase shift between the two emissions, but no  systematic
offset was found as a function of the distance from the nucleus or the side of
the jet (north or south) with respect to the nucleus. We quantified these
results by a thorough correlation coefficient analysis. Similar results (no
correlation or anticorrelation) were obtained when comparing the 8~$\mu$m and
X-ray emissions.

In conclusion, we found that the large-scale jet speed in NGC~4258 (likely
around 1000~km~s$^{-1}$; \citeauthor{cecil92} \citeyear{cecil92}) is not
sufficiently high to destroy dust traced by PAH emission on scales which are resolved by our observations. Future observations with higher resolution may detect dust destruction on smaller scales.

\acknowledgments

We thank Dr. Yuxuan Yang for sharing the reduced Chandra X-ray data with us. We are grateful to Dr. Joe Hora for allowing us to use his custom muxbleed 
correction software. This work is based in part on observations made with the
{\it Spitzer Space Telescope}, which is operated by the Jet Propulsion
Laboratory, California Institute of Technology under a contract with NASA.
Support for this work was provided by NASA through an award issued by
JPL/Caltech.

{\it Facilities:} \facility{Spitzer (IRAC)}

\end{document}